\documentclass{PoS}

\title{Matter swapping between two braneworlds from the equivalence between two-brane
worlds and noncommutative two-sheeted spacetimes}

\ShortTitle{Matter swapping between two branes}

\author{\speaker{Micha\"{e}l Sarrazin}\\
        Department of
Physics,\\ University of Namur (FUNDP),
\\61 rue de Bruxelles, B-5000 Namur, Belgium\\
        E-mail: \email{michael.sarrazin@fundp.ac.be}}

\author{Fabrice Petit\\
        Belgian Ceramic Research
Centre,\\4 avenue du gouverneur Cornez, B-7000 Mons, Belgium\\
       E-mail: \email{f.petit@bcrc.be}}

\abstract{It is shown that a two-brane world made of two
domain walls can be seen as a noncommutative two-sheeted spacetime
under certain assumptions. This equivalence implies a
model-independent phenomenology: Matter swapping between the two
$3$-branes (or sheets) is predicted through fermionic oscillations
induced by magnetic vector potentials. This phenomenon, which might
be experimentally studied, could reveal the existence of extra
dimensions in a new and very affordable way.}

\FullConference{The 2011 Europhysics Conference on High Energy
Physics-HEP 2011,\\
July 21-27, 2011\\
Grenoble, Rh\^one-Alpes France}

\begin{document}

\section{Fermion dynamics in a two-brane world and noncommutative geometries}

Regarding the dynamics of spin-$1/2$ particles, at low energies any
two-brane world related to a domain wall approach [1] is formally
equivalent to a noncommutative two-sheeted spacetime $M_4\times Z_2$
[2]. The demonstration [2] of this result is inspired by quantum
chemistry and the construction of molecular orbitals, here extended
to branes. Let us consider, for instance, a two-brane world made of
two domain walls (two kink-like solitons) on a continuous $M_4\times
R_1$ manifold from:
\begin{equation}
S=\int \left[ -\frac 1{4G^2}\mathcal{F}_{AB}\mathcal{F}^{AB}+\frac
12g^{AB}\left( \partial _A\Phi \right) \left( \partial _B\Phi
\right)
-V(\Phi )+\overline{\Psi }\left( i\Gamma ^A\left( \partial _A+i\mathcal{A}%
_A\right) -\lambda \Phi \right) \Psi \right] \sqrt{g}d^5x  \label{1}
\end{equation}
The continuous real extra dimension $R_1$ can be replaced by an
effective phenomenological discrete two-point space $Z_2$. At each
point along the
discrete extra dimension $Z_2$ there is then a four-dimensional spacetime $%
M_4$ endowed with its own metric field. Both branes/sheets are then
separated by a phenomenological distance $\delta $ which is
inversely proportional to the overlap integral of the
extra-dimensional fermionic wave functions of each $3$-brane over the
fifth dimension $R_1$. Considering the electromagnetic gauge field,
it has been also demonstrated that the
five-dimensional $U(1)$ bulk gauge field is substituted by an effective $%
U(1)\otimes U(1)$ gauge field acting in the $M_4\times Z_2$
spacetime. It is important to stress that the equivalence between
the continuous two-domain wall approaches and the noncommutative
two-sheeted spacetime model is rather general and does not rely for
instance on the domain walls features or on the bulk dimensionality.

The dynamics of a spin-$1/2$ fermion can be then described with a
two-brane Dirac equation [2,3]. The derivative operator is:
$D_\mu =\mathbf{1}_{8\times 8}\partial _\mu $ ($\mu=0,1,2,3$) and$\ D_5=ig\sigma
_2\otimes
\mathbf{1}_{4\times 4}$ with $g=1/\delta$. One can build the Dirac operator defined as ${%
\not{D}=}\Gamma ^ND_N=\Gamma ^\mu D_\mu +\Gamma ^5D_5$ where: $\Gamma ^\mu =%
\mathbf{1}_{2\times 2}\otimes \gamma ^\mu $\ and\ $\Gamma ^5=\sigma
_3\otimes \gamma ^5$. $\gamma ^\mu $ and $\gamma ^5=i\gamma
^0\gamma ^1\gamma ^2\gamma ^3$ are the usual Dirac matrices and $\sigma _k$ (%
$k=1,2,3$) the Pauli matrices. By introducing a general mass term
$M$, a two-brane Dirac equation is then derived:
\begin{equation}
\left( {i{\not{D}}_A-M}\right) \Psi =\left(
\begin{array}{cc}
i\gamma ^\mu (\partial _\mu +iqA_\mu ^{+})-m & ig\gamma
^5-im_r+i\gamma ^5\Upsilon \\
ig\gamma ^5+im_r+i\gamma ^5\overline{\Upsilon } & i\gamma ^\mu
(\partial _\mu +iqA_\mu ^{-})-m
\end{array}
\right) \left(
\begin{array}{c}
\psi _{+} \\
\psi _{-}
\end{array}
\right) =0  \label{2}
\end{equation}
with $\psi _{\pm}$ the wave functions in the branes $(\pm)$ and $\overline{%
\Upsilon }=\gamma ^0\Upsilon ^{\dagger }\gamma ^0$. The off-diagonal
mass term $m_r$ can be justified from a two-brane(-domain wall)
structure of the Universe. The electromagnetic field $\left\{
A_\mu ^{\pm },\Upsilon
\right\} $ is introduced through: ${\not{D}}%
_A\rightarrow {\not{D}}+\not{A}$, according to the $U(1)\otimes U(1)$
gauge group. Assuming that the electromagnetic field
of a brane couples only with the particles belonging to the same
brane, we set $\Upsilon \sim 0$.

\section{Two-brane Pauli equation and model-independent phenomenology}

The non-relativistic limit of the Dirac equation is derived to
obtain a two-brane Pauli equation [2,3]:
$i\hbar \partial_{t}\Psi =\left\{ \mathbf{H}_0+\mathbf{H}_{cm}+\mathbf{...}\right\}
\Psi$, with $\mathbf{H}_0=diag(\mathbf{H}_{+},\mathbf{H}_{-})$, where $\mathbf{H%
}_{\pm }$ are simply the usual four-dimensional Pauli Hamiltonian
expressed in each branes. Moreover, a new fundamental coupling term
appears [2,3] specific to the two-brane world:
\begin{eqnarray}
\mathbf{H}_{cm}=igg_s\mu \frac 12\left(
\begin{array}{cc}
0 & -\mathbf{\sigma \cdot }\left\{ \mathbf{A}_{+}-\mathbf{A}_{-}\right\} \\
\mathbf{\sigma \cdot }\left\{ \mathbf{A}_{+}-\mathbf{A}_{-}\right\}
& 0
\end{array}
\right)  \label{4}
\end{eqnarray}
$\mathbf{A}_{\pm}$ are the magnetic vector
potentials in
the branes $(\pm)$. $%
g_s\mu $ is the magnetic moment of the particle. This specific term
induces a coupling between the two branes through the magnetic
vector potentials of each brane and the fermionic magnetic moment.

The two-brane Pauli equation supports resonant solutions [4].
Let us consider a neutron under the influence of a rotative
magnetic vector potential $\mathbf{A}_p$ (with an angular frequency
$\omega $) localized in our brane. The probability to find the
neutron in the second brane is [4]:
\begin{equation}
P(t)=\frac{4\Omega _p^2}{(\Omega _0-\omega )^2+4\Omega _p^2}\sin
^2\left( (1/2)\sqrt{(\Omega _0-\omega )^2+4\Omega _p^2}t\right)
\label{5}
\end{equation}
where $\Omega _p=gg_s\mu A_p/(2\hbar )$ and $\Omega _0=(V_{+}-V_{-})/\hbar $%
, which defines the interactions of the particle with its environment ($%
V_{\pm }$ are the potential energies of the particle in each brane). When $%
\omega =\Omega _0$, the particle then resonantly oscillates between the branes.

To investigate this matter swapping effect, a possibility would be to study the population of a stored ultracold neutron gas.
The disappearance of neutrons by the presently discussed mechanism could be observed by counting the remaining neutrons in a vessel.
Two kind of experimental devices could be used to that end. Some papers [5] suggest the existence of an astrophysical ambient magnetic vector potential from astrophysical magnetic fields. Maybe, this ambient field could be responsible for non-resonant swapping, which could be soon be tested by teams working with ultracold neutrons [2,3]. Also, a resonant experiment [4] involving pulsed monochromatic electromagnetic radiation could be considered as an artificial means to produce this matter swapping.

\section{Conclusions and outlooks}

A universe which contains at least two branes can be modeled by a $M_4\times Z_2$ two-sheeted spacetime in the formalism
of the noncommutative geometry. The dynamics of fermions in a two-brane world can then be studied independently
of the domain wall formalism. A new effect, which corresponds to an exchange of fermionic matter between the two
braneworlds may be revealed by using convenient magnetic vector potentials. This matter swapping might be investigated by using the current technology. A work is in progress to compare the predictions of the model with already published experimental data.
In a forthcoming study, it will be shown that this presently discussed effect could also be included in the superstring formalism. The study of the dynamics induced by the variations of the coupling constant $g$ is scheduled.

\end{document}